\documentclass{PoS}

\title{Follow-up of X-ray transients detected by {\it SWIFT} with {\it COLORES} using the {\it BOOTES} network}

\ShortTitle{{\it SWIFT} transients observed with {\it COLORES}}

\author{M. D. Caballero-Garcia$^a$, M. Jel\'{\i}nek$^b$, A.~J. Castro-Tirado$^{bc}$, R. Hudec$^{ad}$, R. Cunniffe$^b$, O. Rabaza$^e$ and L. Sabau-Graziati$^f$\\
\\
 \llap{$^a$} Czech Technical University in Prague, Faculty of Electrical Engineering \\ 
             Technick\'a 2, 166 27  Praha 6 (Prague), Czech Republic \\
        E-mail: \email{cabalma1@fel.cvut.cz}                \\
 \llap{$^b$} Instituto de Astrof\'{\i}sica de Andaluc\'{\i}a (IAA-CSIC) \\
             P.O. Box 03004, E-18080, Granada, Spain \\ 
 \llap{$^c$} Unidad Asociada Departamento de Ingenier\'{\i}a de Sistemas y Autom\'atica, E.T.S. de Ingenieros Industriales, Universidad de M\'alaga \\
             M\'alaga, Spain \\
 \llap{$^d$} Astronomical Institute, Academy of Sciences of the Czech Republic \\
             251~65~Ond\v{r}ejov, Czech Republic \\
 \llap{$^e$} \'Area de Ingenier\'{\i}a El\'ectrica, Dpto. de Ingenier\'{\i}a Civil, Univ. de Granada \\
             Granada, Spain \\
 \llap{$^f$} Divisi\'on de Ciencias del Espacio (INTA) \\
             Torrej\'on de Ardoz, Madrid, Spain \\
}

\abstract{ The Burst Observer and Optical Transient Exploring System (BOOTES) is a
network of telescopes that allows the continuous monitoring of transient astrophysical sources. It was
originally devoted to the study of the optical emission from gamma-ray
bursts (GRBs) that occur in the Universe. In this paper we show the initial
results obtained using the spectrograph COLORES (mounted on BOOTES-2),
when observing compact objects of diverse nature.
}

\FullConference{Swift: 10 Years of Discovery,\\
		2-5 December 2014\\
		La Sapienza University, Rome, Italy }

\begin{document}

\section{Introduction}

Compact objects result from the end points of massive stars. When the burning of the star runs out and if the star is massive enough (superseding certain stability
limits) then collapse occurs. In the case of stars with a core mass of $<1.4\,{\rm M}_{\odot}$ a white dwarf (WD) of electron-degenerate matter is formed that, if not exceeding this 
limit, will be radiating energy (and cooling down) in a (more or less) stable way for ${\ge}10^9\,{\rm Gyr}$. In the case of more massive stars, the evolution is more complicated, 
since the core of electron-degenerate-matter is not able to sustain gravitational forces and collapse occurs. Gravitational sources can stop the fatal end only for stars with 
a core $<3\,{\rm M}_{\odot}$, thus forming a neutron star (NS), i.e. core of neutron-degenerated matter. Above this mass limit, there is no force in Nature able to stop the gravitational 
collapse and a black-hole (BH) is formed.

The astrophysical study of compact objects is only possible due to the emission of their associated accretion (either in the form of discs or streams of matter -- the latter in the 
case of the polar cataclysmic variables) in X-rays. For this accretion disc to exist either a donor star (in the case
of binary systems) or material (in the case of Active Galactic Nuclei -- AGN) has to be present. In principle, an isolated compact object can not emit detectable radiation
by current instrumentation. Nevertheless, we also refer to the recent advances that have shown that isolated neutron stars (the so called ``magnetars'') emit huge amounts of 
hard X-rays and/or ${\gamma}$-rays detected by current instrumentation (\cite{kou99}). Therefore, observations in the X-ray regime of compact objects constitute the most important
tool in order to understand the physical properties of matter in the very inner regions around them. Although the classification of X-ray emitting compact objects is 
rather complex (\cite{camen07}) we will be referring here to three basic types: 1) cataclysmic variables; 2) accreting pulsars and 3) black hole X-ray binaries (i.e. composed 
by a WD, NS and BH plus a donor star for the cases 1), 2) and 3), respectively).

The X-ray emission from compact objects is highly variable (due to the small-size scales of the emitting regions involved). The main goal of the X-ray {\it Swift} satellite (\cite{gehrels04})
is the detection of Gamma-Ray Bursts (GRBs) due to its high angular field of view and rapid (time) reaction. Nevertheless, it has also been revealed to be a fantastic tool for the study 
of the X-ray emission from compact objects (with often an unpredictable emission behaviour). Observations in the optical are performed by big and medium-sized telescopes on Earth. The 
former are not suitable for performing the rapid follow-up needed for the study of optical transients (as we will explain hereafter). These transients events are typically of 
short duration (from fractions of a second to a few days), because the physical processes that originate them are of limited duration/spatial extent. Robotic smaller telescopes are
very well suited for performing such studies. This is due to several factors: their observing flexibility, their rapid response and
slew times and the fact that they can be located worldwide working remotely (therefore allowing continuous monitoring). Of course, additional
observations might be triggered after the transient has been detected with large X-ray/Optical Observatories. In this way we can perform
deep studies on the nature of these sources.

\subsection{The Burst Optical Observer and Transient Exploring System and its Spectrographs}

{\it BOOTES} (acronym of the Burst Observer and Optical Transient Exploring System) is a world-wide network of robotic telescopes. It
was originally designed from a Spanish-Czech collaboration that started in 1998 (\cite{castro99,castro12}). The telescopes are located
at Huelva ({\it BOOTES}--1), M\'alaga ({\it BOOTES}--2), Granada,
Auckland ({\it BOOTES}--3) and Yunnan ({\it BOOTES}--4), located at Spain, New Zealand and China, respectively. There are plans of extending
this network even further (Mexico, South Africa, Chile,...). These
telescopes are medium-sized (${\rm D}=30-60$\,cm), autonomous and very versatile. They are very well suited for the continuous study of the fast variability from
sources of astrophysical origin (GRBs and Optical Transients -- hereafter OTs -- ).

Currently two spectrographs are built and working properly in the network at M\'alaga and Granada (in the optical and infra-red, respectively). In
Sec.~\ref{results} we will show preliminary results obtained so far with {\it COLORES} at {\it BOOTES}--2 in the field of compact objects.

\subsection{{\it COLORES}}

{\it COLORES} stands for Compact Low Resolution Spectrograph (\cite{rabaza14}). It is a spectrograph designed to be lightweight enough to be carried by the high-speed robotic
telescope 60\,cm ({\it BOOTES}--2). It works in the wavelength range of ($3\,800-11\,500$)\,{\AA} and has a spectral resolution of ($15-60$)\,{\AA}. The
primary scientific target of the spectrograph is a prompt GRB follow-up, particularly the estimation of redshift.

{\it COLORES} is a multi-mode instrument that can switch from imaging a field (target selection and precise pointing) to spectroscopy by rotating wheel-mounted
grisms, slits and filters. The filters and the grisms (only one is mounted at the moment) are located in standard filter wheels and the optical design is
comprised of a four-element refractive collimator and an identical four-element refractive camera. As a spectroscope, the instrument can use different
slits to match the atmospheric seeing, and different grisms in order to select the spectral resolution according to the need of the observation.

The current detector is a $1\,024{\times}1\,024$ pixels device, with 13 micron pixels. The telescope is a rapid and lightweight design, and a low instrument weight was
a significant constraint in the design as well as the need to be automatic and autonomous. For further details on description, operation and working with
{\it COLORES} we refer to M. Jelinek PhD thesis (\cite{jelinek14} and references therein).

\section{Scientific results obtained using {\it BOOTES}-2/COLORES}  \label{results}

In the following we present some important scientific results obtained using {\it BOOTES}-2 and its low-resolution spectrograph {\it COLORES}, recently obtained  
(we refer to \cite{caballe14a} for a summary of previously obtained results in the field of OTs).

\subsection{The accreting pulsar A~0535+262}

The evolution of the ${\rm H}_{\alpha}$ equivalent width (EW) of the Be/X-ray binary system A~0535+262 (also called HD~245770) 
was reported (\cite{camero14b}), using observations performed with the spectrograph {\it COLORES} at the 0.6\,m telescope {\it BOOTES}--2 (M\'alaga, 
Spain) on 2015-01-27 at 22:05:31.736 UTC (MJD~57049.920), and with the spectrograph located at the 51\,cm telescope of the Observatorio de 
Aras de los Olmos of the University of Valencia on 2015-01-29 at 01:00:00 UTC (MJD~57051.042).

The results indicate that the ${\rm H}_{\alpha}$ line (in emission) had an EW of the order of ($-6.6{\pm}0.9$)\,{\AA} (MJD~57049.920) 
and ($-7.9{\pm}0.2$)\,{\AA} (MJD~57051.042). This means that the size of the Be disk had decreased to half of the value measured on 
2014 March 19 (\cite{camero14a}), being similar to the estimate during quiescence on 2012 March. Recently, this system showed a 
weak enhancement in the optical luminosity (\cite{giovannelli15}), and a weak reawakening of X-ray pulsar's activity was thus expected. MAXI 
detected this source on 2015 January 29 (\cite{nakajima15}), and the Swift/BAT monitoring web site was detecting its rise with a flux of 
$570{\pm}20$\,mCrab (MJD~57057).

The behavior of this system might resemble the giant outburst of 2011 in which the EW of the ${\rm H}_{\alpha}$ line was initially of the 
order of $-9.5$\,{\AA} (\cite{camero12}).

\begin{figure*}
\centering
\includegraphics[bb=0 0 612 792,angle=270,width=0.8\linewidth]{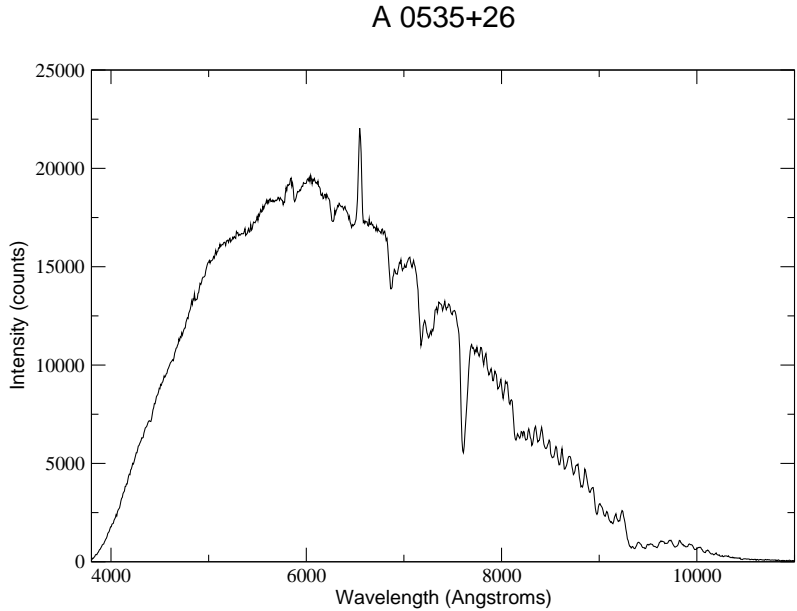}
\caption{Optical spectrum in the range (3\,800-11\,000)\,${\AA}$ from A~0535+26, obtained with {\it BOOTES}-2/COLORES on $16^{\rm th}$ Oct. (2014) at 01:37 UT.}
\label{fig1}
\end{figure*}

\subsection{The accreting black hole SS~433}

In light of the recent extreme outburst of SS 433 (\cite{goranskij14,charbonnel14}) and follow-up by {\it Swift} (\cite{sokolovsky14}) the 
0.6\,m TELMA robotic telescope at the {\it BOOTES}-2 (+{\it COLORES}) astronomical station in M\'alaga (Spain), obtained optical ($4\,000-9\,000$)\,{\AA} 
spectra starting at 2014-08-01 and ending at 2014-08-07, as reported (\cite{caballe14b}). The most prominent feature of the spectra was the variable ${\rm H}_{\alpha}$ line. There 
was also a variable emission line at 7\,100\,{\AA} that might correspond to a 300\,{\AA} shifted ${\rm H}_{\alpha}$ P-fund line 
(7\,460\,\,{\AA}), previously reported in the literature (\cite{fuchs06}).

\begin{figure*}
\centering
\includegraphics[bb=0 0 612 792,angle=270,width=0.8\linewidth]{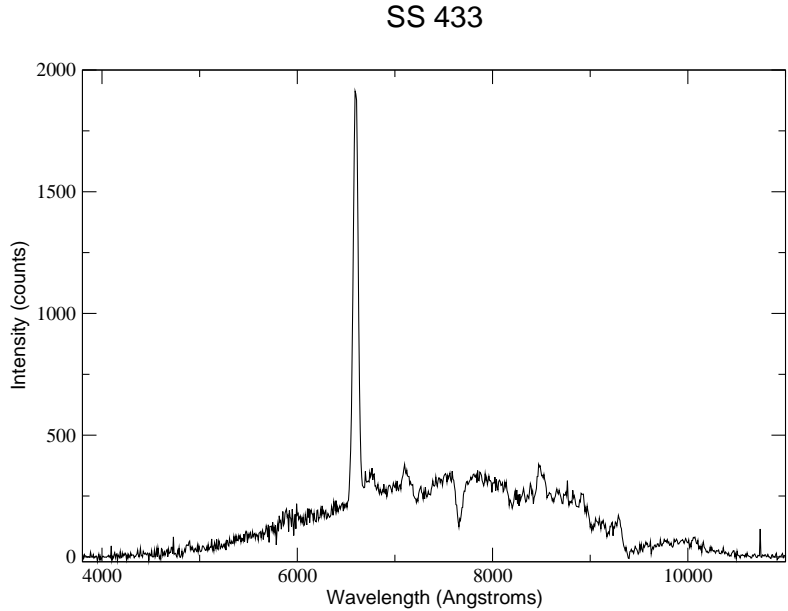}
\caption{Optical spectrum in the range (3\,800-11\,000)\,${\AA}$ from SS~433, obtained with {\it BOOTES}-2/COLORES on $2^{\rm nd}$ Aug. (2014) at 02:33 UT.}
\label{fig1}
\end{figure*}

\section{Discussion and conclusions}  

The era of OTs is about to start. Since the beginning of the modern times the big telescopes have been the only
resource for astronomers to study astrophysical sources. In spite of constituting the best tool for deep studies of individual
targets, they are not properly suited for the discovery of optical transient sources. Their big size limits the speed they can
cover the entire sky and the time for overheads might be longer than that for medium-sized telescopes. Many factors make them difficult to
fully automatize and indeed currently none is completely robotic and autonomous. Medium-sized telescopes (i.e. ${\rm D}{\leq}1$\,m) can be
much quicker moving from target to target and time overheads are usually very small. Therefore, robotic medium-sized telescopes
are currently the best ones for the follow-up and studies of the long-term variability of the astrophysical transient sources.

{\it BOOTES}-2 constitutes one step forward with respect to other robotic telescopes. This is because
it is one (of the very few) robotic telescopes with a spectrograph mounted on it. It has demonstrated to work properly and efficiently and in this 
paper we give a first glimpse on the results obtained from its observations of compact objects.
Since compact objects are transient sources variable in the sky (i.e. OTs) the {\it Swift} satellite constitutes a precious tool for detecting
and following them. When an OT has been detected by {\it Swift} at X-rays, 
an alert is created to the scientific community (often through an {\it Astronomer Telegram} or the {\it Gamma-ray Coordinates Network}). In such a 
case the {\it BOOTES} telescopes start the observation, re-observing the target every time it is visible during the following nights.

Apart from the intensive campaign of follow-up of GRBs performed by the {\it BOOTES} network (${\approx}100$ GRBs have been observed so far), {\it BOOTES}-2
and its spectrograph (COLORES) are also providing excellent results in the field of compact objects. In this paper we mention a few cases, obtained during the last
2\,years (since the spectrograph was mounted on the telescope). But this is only the beginning and we look forward to follow
many compact objects for understanding better their physical properties and may be also to discover new kinds of compact objects (and/or systems
containing a compact object).

\acknowledgments
MCG is supported by the European social fund within the framework of realizing the project "Support of inter-sectoral mobility and 
quality enhancement of research teams at Czech Technical University in Prague“, CZ.1.07/2.3.00/30.0034.

\end{document}